\documentclass[prl,aps,twocolumn,superscriptaddress,showpacs,floatfix]{revtex4-1}
\usepackage[T1]{fontenc}
\usepackage{lmodern}
\usepackage{amsmath,amsfonts}
\usepackage{graphicx}
\newcommand{\figref}[1]{Fig.~\ref{fig:#1}}
\begin{document}
\title{Observation of Defect States in $\mathcal{PT}$-Symmetric Optical Lattices}

\author{Alois Regensburger$^\dagger$}%
\affiliation{Institute of Optics, Information and Photonics, SAOT, University of Erlangen-Nuernberg, 91058 Erlangen, Germany}
\author{Mohammad-Ali Miri$^\dagger$}%
\affiliation{CREOL/College of Optics, University of Central Florida, Orlando, Florida 32816, USA}%
\author{Christoph Bersch}%
\affiliation{Institute of Optics, Information and Photonics, SAOT, University of Erlangen-Nuernberg, 91058 Erlangen, Germany}
\author{Jakob N\"ager}%
\affiliation{Institute of Optics, Information and Photonics, SAOT, University of Erlangen-Nuernberg, 91058 Erlangen, Germany}
\author{Georgy Onishchukov}%
\affiliation{Max Planck Institute for the Science of Light, 91058 Erlangen, Germany}
\author{Demetrios N. Christodoulides}%
\affiliation{CREOL/College of Optics, University of Central Florida, Orlando, Florida 32816, USA}%
\author{Ulf Peschel}%
\email{ulf.peschel@physik.uni-erlangen.de; $^\dagger$: Equal contribution.}%
\affiliation{Institute of Optics, Information and Photonics, SAOT, University of Erlangen-Nuernberg, 91058 Erlangen, Germany}

\begin{abstract}
We provide the first experimental demonstration of defect states in parity-time ($\mathcal{PT}$) symmetric mesh-periodic potentials. Our results indicate that these localized modes can undergo an abrupt phase transition in spite of the fact that they remain localized in a $\mathcal{PT}$-symmetric periodic environment.  Even more intriguing is the possibility of observing a linearly growing radiation emission from such defects provided their eigenvalue is associated with an exceptional point that resides within the continuum part of the spectrum. Localized complex modes existing outside the band-gap regions are also reported along with their evolution dynamics.
\end{abstract}
\pacs{42.81.Qb, 
  42.25.Bs 	
  11.30.Er	
}

\maketitle 

Defects play a crucial role in determining the physical and chemical properties of solids \cite{Ashcroft1976}. In semiconductor crystals, the presence of defects leads to both bulk and surface electronic states that ultimately affect charge transport processes. When analyzed from the viewpoint of their corresponding band structure, such localized quantum eigenstates are known to reside within the forbidden energy gaps. In optics, similar effects are also possible in photonic crystal arrangements which have been so far exploited to realize high quality dielectric waveguides and cavity resonators \cite{Mekis1996, Yablonovitch1991,Trompeter2003}. In most cases, such defect modes have been investigated within the context of Hermitian systems. Yet, much less is known about the physics and properties of defects in non-Hermitian periodic configurations where the vector space is no longer orthogonal but is instead skewed. In the optical domain, non-hermiticity can be readily introduced through either amplification or loss. Such arrangements include for example defect mode lasers in photonic band-gap crystals \cite{Painter1999}, photonic crystal fiber amplifiers \cite{Knight1998} and semiconductor distributed-feedback lasers \cite{Haus1984}.  The spectrum of these latter systems is in general complex, allowing only some of the modes to enjoy amplification.

Lately, the notion of parity-time ($\mathcal{PT}$) symmetry has been introduced in optics as a new paradigm to mold the flow of light \cite{Makris2008, *Musslimani2008, El-Ganainy2007, *Miri2012b}. This idea, which originated within the context of quantum field theories \cite{Bender1998, *Bender2002, *Bender2007, Mostafazadeh2002, *Levai2000, *Ahmed2001}, has led to new strategies in achieving a harmonic interplay between optical gain and loss. In general, a necessary (but not sufficient) condition for an optical structure to be $\mathcal{PT}$-symmetric is that its complex refractive index distribution satisfies the condition $n(x) = n^*(-x)$, in which case the real part of the index profile is expected to be symmetric in space while the imaginary component (gain-loss) is antisymmetric. Optical systems endowed with this symmetry are known to exhibit altogether real spectra. $\mathcal{PT}$ symmetry can lead to unusual and previously unattainable light propagation features \cite{Regensburger2012, Miri2012a, Joglekar2010, Graefe2011, Ruter2010, *Guo2009, Klaiman2008, Midya2010, Ramezani2010, *Bendix2009, Chong2011, *Liertzer2012, Schomerus2010, Miri2012b, Longhi2009, *Longhi2010, Lin2011, *Kulishov2005, *Feng2012, Miroshnichenko2011, *Suchkov2012}. These include double refraction and band merging \cite{Makris2008, *Musslimani2008,Regensburger2012}, abrupt phase transitions and power oscillations \cite{Ruter2010, *Guo2009, Klaiman2008}, unidirectional invisiblity \cite{Lin2011, *Kulishov2005, *Feng2012} and non-reciprocal propagation \cite{Ramezani2010, *Bendix2009}, as well as coexistence of coherent lasing-absorbing modes and mode selection in $\mathcal{PT}$-symmetric lasers \cite{Chong2011, *Liertzer2012, Schomerus2010, Miri2012b}. Quite recently, light transport in large-scale temporal $\mathcal{PT}$-symmetric mesh lattices has been reported \cite{Regensburger2012, Miri2012a}. Given that the band structure of a $\mathcal{PT}$-symmetric lattice can be entirely real, one could ask in what fundamental ways the properties of a defect state will be altered in such a pseudo-Hermitian environment.

\begin{figure}
  \includegraphics[width=0.90\linewidth]{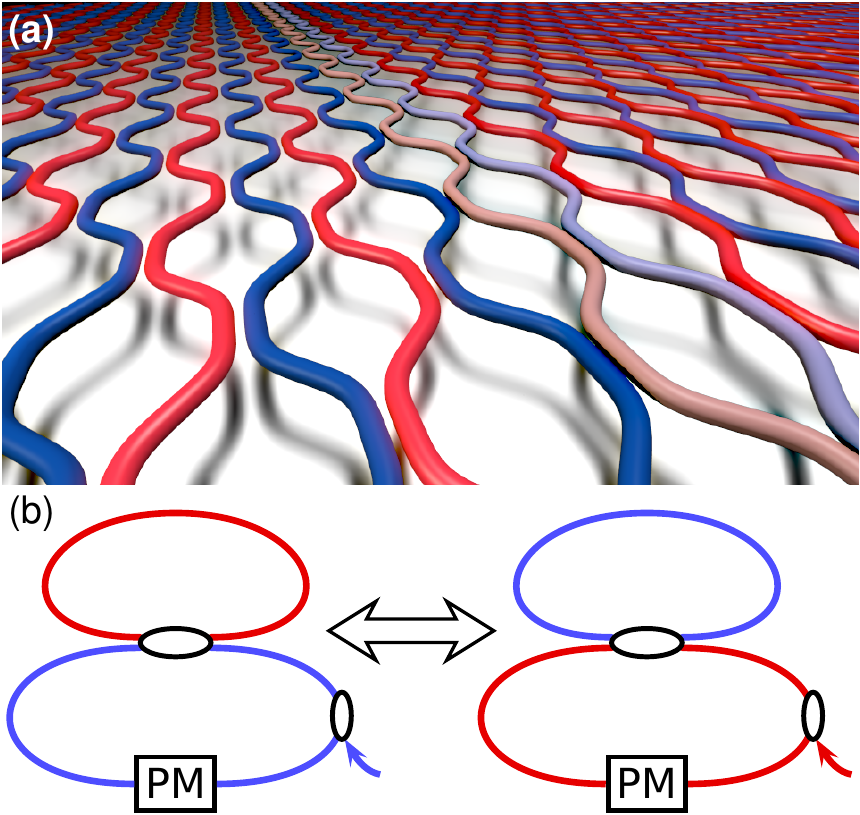}
  \caption{(Color online). (a) Equivalent spatial waveguide mesh lattice corresponding to the temporal fiber loop scheme used in the experiments shown in (b). Red and blue waveguides indicate balanced regions of gain and loss while transverse 50\% coupling takes place where waveguides come close.  A pair of waveguides with a different gain-loss contrast and/or phase shift acts as a defect. (b) Coupled fiber loops used in the experiment. PM:phase modulator.}
  \label{fig:1}
\end{figure}

In this Letter, we report the first experimental observation of defect states in $\mathcal{PT}$-symmetric lattices [\figref{1}(a)]. We demonstrate the transition from stable to exponentially growing bound modes while their localization properties remain preserved. Furthermore, we show that for a defect state in its broken symmetry regime, the corresponding eigenvalue does not necessarily have to reside within the band gap region---as in conventional Hermitian periodic structures. Finally, at $\mathcal{PT}$ threshold, we observe a stable parity-time symmetric defect mode that constantly emits coherent radiation to the surrounding lattice at a linear growth rate. 

The experimental setup in \figref{1}(b) \cite{Regensburger2012, Suppl,Regensburger2011} consists of two loops of optical fiber which are connected by a 50:50 coupler. A length difference of  $\Delta L$ between the loops enables a temporal advancement or delay of the light pulses in every round trip \cite{Schreiber2010, *Schreiber2012, Regensburger2011}. The required refractive index distribution in this arrangement is realized using a phase modulator in one of the loops. Additionally, the antisymmetric imaginary part of the photonic potential is implemented by temporally switching gain and loss between the two loops. A comprehensive technical documentation is provided in the Supplemental Material \cite{Suppl}. One can show that the resulting pulse dynamics in this system is governed by the following difference equations \cite{Regensburger2012, Regensburger2011}: 
\begin{align}
   u_n^{m+1} & = \frac{1}{\sqrt{2}}\,\widetilde{G}(n+1)^{\frac{1}{2}(-1)^m}
     (u_{n+1}^m + iv_{n+1}^m)\\\nonumber
   v_n^{m+1} & =
\frac{1}{\sqrt{2}}\,\widetilde{G}(n)^{-\frac{1}{2}(-1)^{m}}(iu_{n-1}^m +
v_{n-1}^m) \text{e}^{i\widetilde{\varphi}(n)}.
\end{align}
Here, $u_n^m$ and $v_n^m$ are the amplitudes of light pulses circulating in the short and long loop, respectively. $m$ corresponds to the number of round trips and $n$ accounts for the transverse temporal position of a pulse. As indicated in Ref. \cite{Suppl}, Eqs.~(1) can be transformed into standard form associated with mesh lattices \cite{Miri2012a}. 

The phase potential $\widetilde{\varphi}(n) =
\widetilde{\varphi}_p(n) + \widetilde{\varphi}_d(n)$ consists of a periodic part $\widetilde{\varphi}_p = \left\{\begin{array}{@{}l@{}} 
+\varphi_p\text{ for }\text{mod}\,(n, 4) = 0,1\\
     -\varphi_p\text{ for }\text{mod}\,(n, 4) = 2,3\end{array}\right.$, and the phase defect $\widetilde{\varphi}_d(n)$ which takes the value $\varphi_d$ for $n$ within the defect and is 0 elsewhere.

In our setup, from one round trip to another, gain and loss alternate between the loops. In general, the gain factor $\widetilde{G}(n) = G_p + \widetilde{G}_d(n)$ itself can depend on $n$: $\widetilde{G}_d(n)$ takes the value $G_d$ for $n$ inside the defect and is 0 everywhere else, thus creating a defect in the imaginary part of the effective potential. In the experiment, the same transverse profiles for the gain and phase potential are used for every loop round trip $m$.
\begin{figure}
  \includegraphics[width=\linewidth]{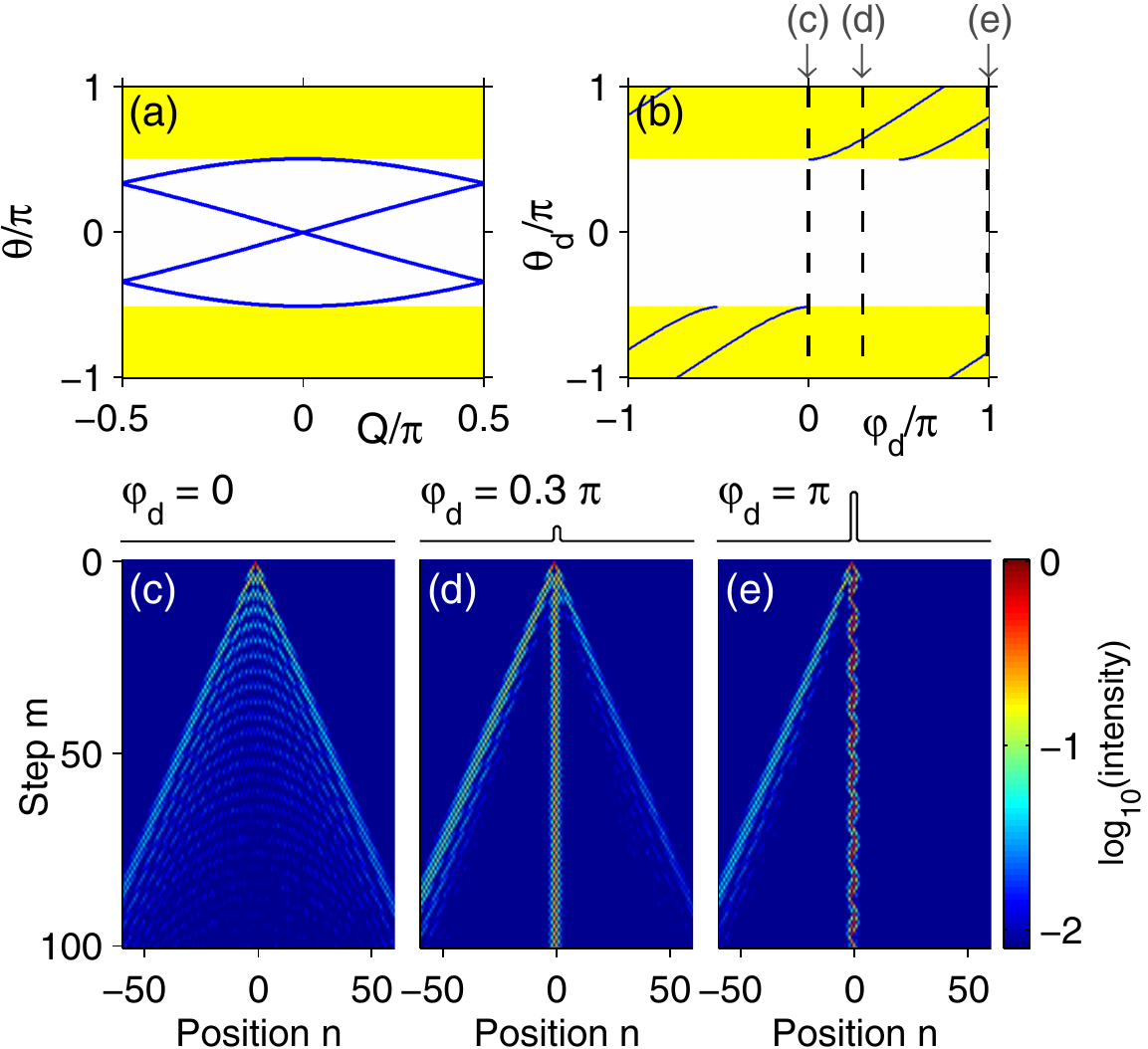}
  \caption{(Color online). Phase defect in a passive mesh lattice. (a) The band structure of the empty lattice ($G_p=1$,
    $\varphi_p=0$) relates the transverse wave number $Q$ and the propagation constant $\theta$. The yellow regions denote band gaps \cite{Miri2012a}. (b) Dispersion curves of the defect modes as a function of the defect phase $\varphi_d$. Experimental results for the evolution of a single pulse in this lattice with a phase defect of strength $\varphi_d$ where (c) $\varphi_d=0$ (no defect), (d) $\varphi_d=0.3\pi$, (e) $\varphi_d=\pi$. The total phase potential $\widetilde{\varphi}(n)$ is also indicated for each case. In all cases only data from the short loop ($u_n^m$) is displayed~\cite{Suppl}.}
  \label{fig:2}
\end{figure}

Before introducing defects in a $\mathcal{PT}$-symmetric optical mesh lattice, we first investigate the corresponding passive Hermitian system \cite{Wojcik2012}. Consider an elemental phase defect $\varphi_d$ at positions $n=\{0,1\}$ in a passive environment ($G_p=1$) where no background potential is present ($\varphi_p=0$). In the absence of a defect, the band structure \cite{Miri2012a} has two connected bands, with the photonic band gaps positioned above and below [\figref{2}(a)] in its reduced Brillouin zone. This can be obtained via a plane wave ansatz of the form $e^{iQn/4}e^{i\theta m/2}$, where $Q$ and $\theta$ represent the transverse Bloch momentum and propagation constants respectively. Such calculations are explained in detail in Refs.~\cite{Miri2012a,Suppl}. The propagation of light pulses in this configuration is classically analogous \cite{Regensburger2011} to a quantum walk \cite{Schreiber2010, *Schreiber2012, Wojcik2012, Bouwmeester1999, Sansoni2012}, as can be seen in \figref{2}(c). Depending on the defect phase $\varphi_d$, either one or two localized modes can exist inside the band gap [\figref{2}(b)]. In our experiment, we inject a single pulse into the long loop which is then monitored during propagation. A convolution of this delta-like impulse with the localized defect states as well as any continuum modes within the bands determines the weights of mode excitation. In the Hermitian system, the strength of each mode remains invariant during propagation. As defect modes are indeed excited, we observe a clear localization of light along the phase defect [Figs.~\ref{fig:2}(d,e)]. For $\varphi_d = \pi$, an oscillatory intensity pattern reveals the presence of two defect modes which continuously interfere with each other. The period of this beating in \figref{2}(e) is close to 10 steps in $m$, which is compatible with the difference between the two associated propagation constants $\Delta\theta_d\approx0.4\pi=\frac{2}{10}2\pi$.

\begin{figure}
  \includegraphics[width=\linewidth]{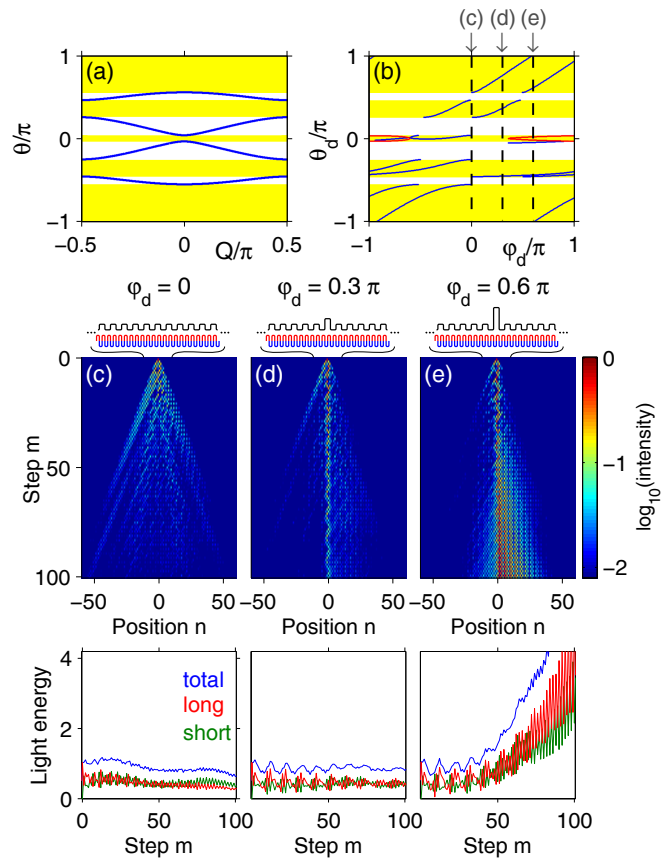}
  \caption{(Color online). Phase defect in a $\mathcal{PT}$-synthetic lattice. (a) Real band structure of the background $\mathcal{PT}$ lattice (below treshold) with a gain-loss $G_p=1.3$ and a phase potential $\varphi_p=0.2\pi$. (b) Dispersion diagram of defect states as a function of defect phase $\varphi_d$. Here, real and imaginary parts are shown in blue and red respectively. (c) Measurement of single pulse evolution when $\varphi_d=0$ (no defect) and for (d) $\varphi_d=0.3\pi$. In this last case, bound modes with real propagation constants $\theta_d$ are observed. (e) Increasing $\varphi_d$ to $0.6\pi$ brings a pair of bound modes above the $\mathcal{PT}$ threshold. The real ($\widetilde{\varphi}(n)$) and imaginary ($\widetilde{G}(n)$) parts of the $\mathcal{PT}$ potentials as well as the evolution of the total power are indicated.}
  \label{fig:3}
\end{figure}

A more complex behavior arises when the same defect is introduced into a $\mathcal{PT}$-symmetric mesh lattice having a balanced optical gain-loss profile ($G_p=1.3$) and a periodic phase (refractive index) potential ($\varphi_p=0.2\pi$). Given that the periodic lattice parameters $G_p$ and $\varphi_p$ were chosen to be below the $\mathcal{PT}$ threshold of this lattice ($\ln{(G_p)}<\cosh^{-1}{\left[2\cos{(\varphi_p)}-\sqrt{\cos{(2\varphi_p)}}\right]}$) \cite{Regensburger2012, Miri2012a}, in the absence of a defect, the band structure of this non-Hermitian system is entirely real [\figref{3}(a,c)]. In this manner none of the Floquet-Bloch modes can grow exponentially. In what follows, we will only consider defects that satisfy the $\mathcal{PT}$ condition of $n(x) = n^*(-x)$ in the entire system (lattice plus defect). 

We now introduce a similar phase defect into a $\mathcal{PT}$ lattice while the imaginary part of the combined $\mathcal{PT}$ potential remains completely periodic. As predicted by our theoretical results,   [\figref{3}(b)], a transition between almost stable localized modes with real propagation constants and exponentially growing bound states with complex eigenvalues is possible, which is in fact observed [\figref{3}(d,e)] in our experiments. Indeed, in the latter case, a pair of defect modes with broken $\mathcal{PT}$ symmetry (complex eigenvalues) emerges. Remarkably, this transition happens when increasing the defect potential $\varphi_d$ while the gain-loss $G_p$ and background potential $\varphi_p$ are kept constant, see \figref{3}(a). This is counterintuitive given that for a homogeneous lattice, such an increase in the optical potential's real part typically leads to stabilization \cite{Regensburger2012}.

Thus far, the propagation constants $\theta_d$ of the bound defect modes were all found to lie within the band gap of the periodic structure, thus prohibiting any coupling to phase-matched free propagating radiation modes. Note that the only way bound states can exist inside the continuum of bands (in Hermitian systems) is when they are totally decoupled from their surroundings by virtue of some special symmetry \cite{Plotnik2011, *Molina2012, Friedrich1985}.  

As we will see, in stark contrast to Hermitian systems, in the case of $\mathcal{PT}$ lattices defect states with complex eigenvalues can also appear within the band continuum. If a defect possesses an imaginary $G_d$ component (in addition to the real part $\varphi_d$) which differs from the periodic gain-loss $G_p$ profile, then it is possible to establish localized modes having propagation constants whose real parts $\text{Re}(\theta_d)$ are located inside the bands. In this regime, the inherent gain of the system compensates for light leaking away (because of phase-matching) into lattice radiation modes. These defect modes have a non-zero imaginary part $\text{Im}(\theta_d)$ that reveals itself through exponential growth or decay. Despite the fact that their coupling to the continuum is not inhibited, these $\mathcal{PT}$ localized states still decay exponentially on both sides. Essentially, this exponential localization is a direct outcome of the exponential increase a defect mode experiences in time.

\begin{figure}
  \includegraphics[width=\linewidth]{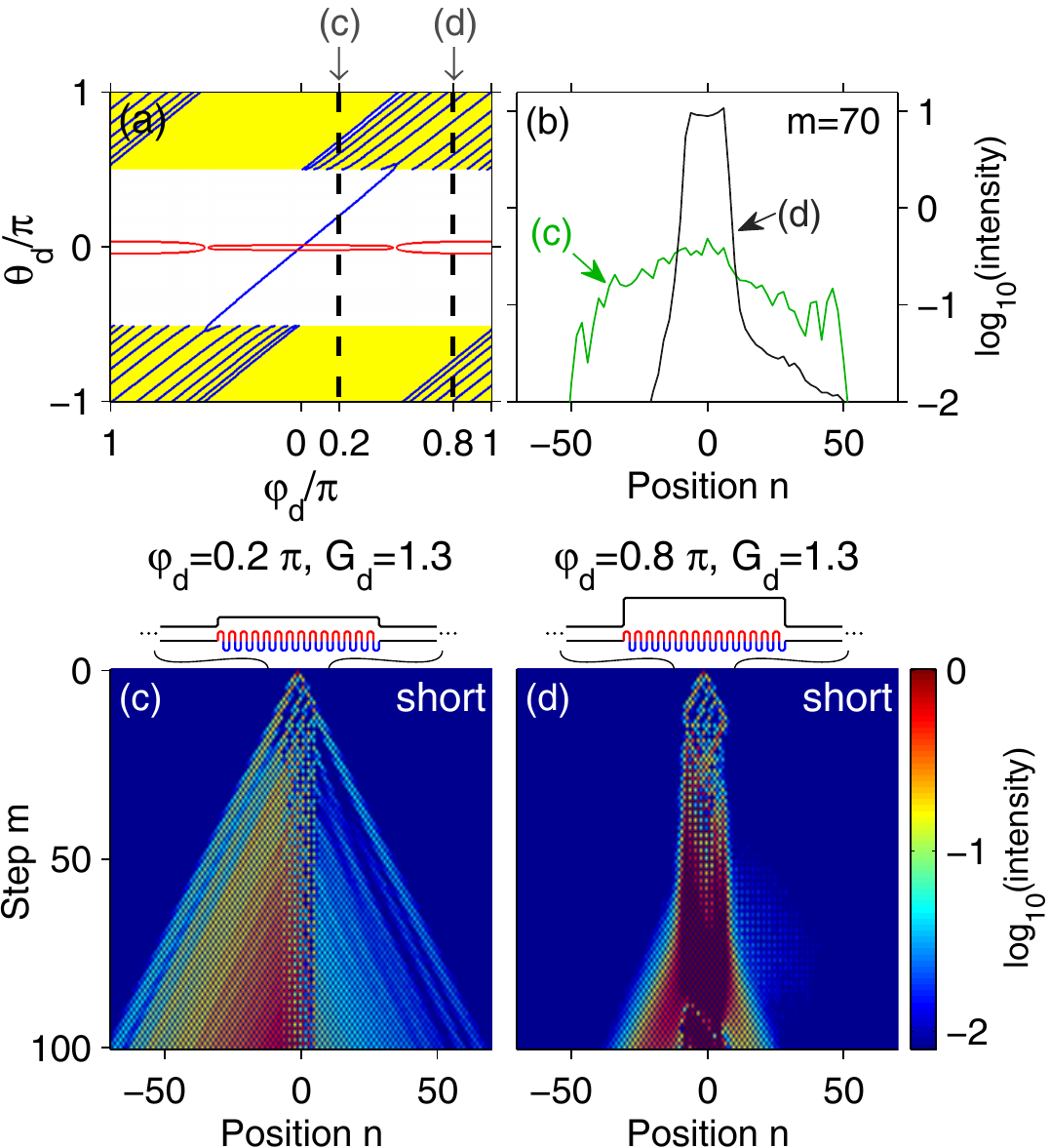}
  \caption{(Color online). Defect modes residing in the continuum. (a) Dispersion diagram of a broad $\mathcal{PT}$-symmetric defect when embedded into an empty lattice. Apart from several other defect modes having real eigenvalues and located in the band gap (yellow region), for $\varphi_d<0.5\pi$ this structure also supports a pair of localized modes with complex $\theta_d$, that lie inside the continuum. (b) Pulse intensity profiles after  $m=70$ steps of propagation.  (c) Observation of a growing weakly localized mode in the continuum for $\varphi_d=0.2\pi$. (d) Increasing $\varphi_d$ to $0.8\pi$ brings the mode into the band gap, leading to strong localization.}
  \label{fig:4}
\end{figure}

In order to observe such defect states in optical mesh lattices, we introduce a broad $\mathcal{PT}$-symmetric defect in a background empty lattice ($G_p=1$ and $\varphi_p=0$). The defect region extends over 14 discrete positions $n$ and possesses a gain-loss contrast of $G_d=1.3$ and a defect phase of $\varphi_d=0.2\pi$, see Ref.~\cite{Suppl}. This extended defect allows one to observe these effects at experimentally attainable gain-loss values. According to the dispersion diagram of Fig.~\ref{fig:4}(a), this lattice can in general support several defect modes. In the range of $0<|\varphi_d|<0.5\pi$, a pair of defect states with complex conjugate eigenvalues (one growing while the other one decaying) is found to exist in spite of the fact that their corresponding real part resides inside the band continuum. In our experiment, by choosing $\varphi_d=0.2\pi$ we clearly observe this weakly localized exponentially growing defect state [Figs.~\ref{fig:4}(b,c)]. When increasing the defect's real potential $\varphi_d$ to $0.8\pi$, this pair of complex defect eigenmodes migrates into the band gap thus becoming tightly bound to the defect site [\figref{4}(d)] while it still grows/decays exponentially.

\begin{figure}
  \includegraphics[width=\linewidth]{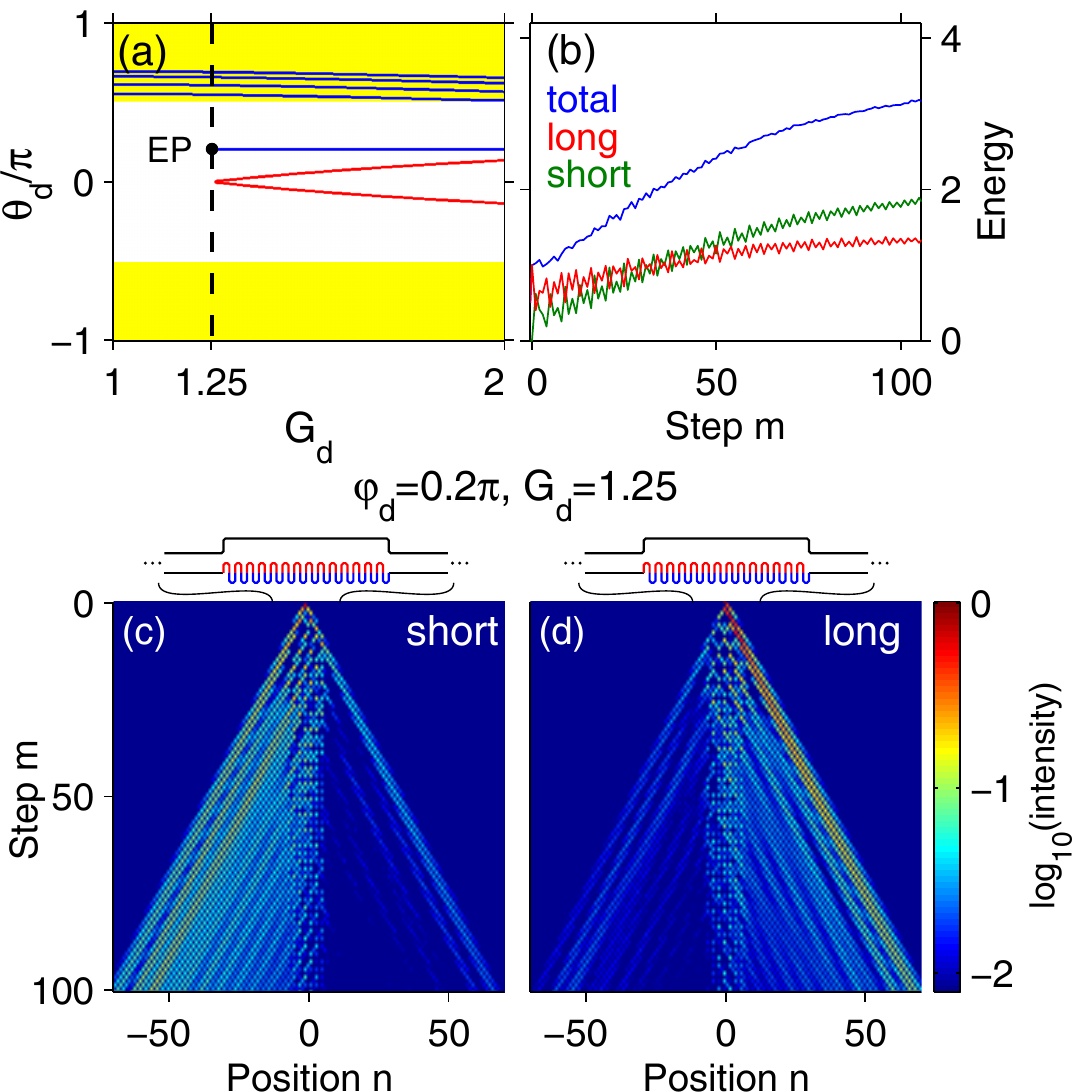}
  \caption{(Color online). A defect mode at its exceptional point (EP). (a)
    Dispersion diagram of defect modes existing in the same structure as in \figref{4} with
    $\varphi_d=0.2\pi$ and as a function of $G_d$. According to this plot $\mathcal{PT}$ threshold occurs at
    $G_d\approx 1.25$. (b) Measured light energy and (c,d) propagation
    in short and long loops at $G_d\approx 1.25$, confirming an
    almost linear growth in total energy and a continuous emission of
    power.}
  \label{fig:5}
\end{figure}

Even more surprising is how in this same structure, the gain-loss coefficient $G_d$ ultimately affects the properties of these complex defect states. According to \figref{5}(a), when $G_d$ decreases (for $\varphi_d=0.2\pi$), the defect eigenvalue spectrum becomes again real and hence the total energy in the lattice remains bounded. However, right at the transition threshold which corresponds to $G_d=1.25$, the mode is no longer exponentially localized and the total light power in the lattice now grows linearly while the rest of the defect eigenvalues are still real [\figref{5}(b)]. In this case, the pulse intensity within the active defect region oscillates around a stable mean value, while the structure constantly emits optical power toward both sides, see Figs.~\ref{fig:5}(c,d). This can be understood from \figref{5}(a) which clearly indicates the presence of an exceptional point within the band. Here a transition occurs between a pair of radiation modes (with real eigenvalues) and two localized complex modes endowed with a finite norm. Therefore, it is possible to create a continuously emitting coherent light source within a photonic lattice by embedding an appropriately designed gain-loss defect, as demonstrated in our experiment [\figref{5}(c)], which indeed confirms a linear growth of total light power [\figref{5}(b)]. Note that a similar behavior occurs in active Fabry-Perot cavities when operated at lasing threshold; in this regime, when gain is exactly equal to the total loss, power remains constant within the cavity while coherent laser light constantly flows toward the outside environment, with the total energy growing linearly \cite{Siegman1986}. 

In conclusion, we have investigated, both theoretically and experimentally, the properties of complex defects in $\mathcal{PT}$-symmetric optical lattices. We have shown that defect modes in such structures can exhibit extraordinary characteristics that are by no means attainable in standard Hermitian systems. Among them is the prospect of $\mathcal{PT}$-symmetry breaking instabilities and the possibility of establishing localized complex defect modes with spectra lying within the band continuum. In such $\mathcal{PT}$-symmetric environments, not only light beams can be trapped within a defect, but can also be controlled at will through a defect parameter--thus altering the respective power emission characteristics to the surrounding regions. Our results may lead to new possibilities in judiciously structuring gain and loss in optical lattices that could in turn be potentially useful in lasing systems and other optical structures and devices. 

\acknowledgments{We acknowledge financial support from DFG Forschergruppe 760, the Cluster of Excellence Engineering of Advanced Materials and the German-Israeli Foundation. This work was also supported by NSF grant ECCS-1128520 and by AFOSR grant FA95501210148. Moreover, we thank M. Wimmer for discussions.}


%

\end{document}


\begin{center}
    \bfseries\Large
    Supplemental material to:

    Observation of Defect States in $\mathcal{PT}$-Symmetric Optical Lattices
  \end{center}
  \vspace*{2mm}
  \begin{center}
    Alois~Regensburger\textsuperscript{1},
    Mohammad-Ali~Miri\textsuperscript{2},
    Christoph~Bersch\textsuperscript{1},
    Jakob~N\"ager\textsuperscript{1},
    Georgy~Onishchukov\textsuperscript{3},
    Demetrios~N.~Christodoulides\textsuperscript{2},
    Ulf~Peschel\textsuperscript{1}
  \end{center}
  \begin{center}
    \textsuperscript{1} \textit{Institute of Optics, Information and Photonics, SAOT,\\ University of Erlangen-Nuernberg, 91058 Erlangen, Germany}

    \textsuperscript{2} \textit{CREOL/College of Optics, University of Central Florida, Orlando, Florida 32816, USA}

    \textsuperscript{3} \textit{Max Planck Institute for the Science of Light, 91058 Erlangen, Germany}

  \end{center}
  \vspace*{5mm}

  \tableofcontents
  \vspace*{5mm}

\section{Evolution equations and equivalence with mesh lattices}
Here we show that the evolution of pulses in the fiber loops [Eqs.~(1)], can be transformed to the standard evolution equations of the optical mesh lattice which is investigated in detail in Ref.~\cite{Miri2012a}. We show this equivalence for the Hermitian case (when no gain or loss is involved). The $\mathcal{PT}$-symmetric case can be explored in a similar fashion. Consider Eqs.~(1) for even round-trips:
\begin{subequations}\label{eqs:1}
\begin{align}
	u_{2k}^{2l+2} & =\frac{1}{\sqrt{2}}\left(u_{2k+1}^{2l+1}+iv_{2k+1}^{2l+1}\right),\\
	v_{2k}^{2l+2} & =\frac{1}{\sqrt{2}}\left(v_{2k-1}^{2l+1}+iu_{2k-1}^{2l+1}\right)e^{i\varphi_{2k}}.
\end{align}
\end{subequations}
After rewritting these same equations for odd round-trips, we obtain: 
\begin{subequations}\label{eqs:2}
\begin{align}
	u_{2k+1}^{2l+1} & =\frac{1}{\sqrt{2}}\left(u_{2k+2}^{2l}+iv_{2k+2}^{2l}\right),\\
	v_{2k+1}^{2l+1} & =\frac{1}{\sqrt{2}}\left(v_{2k}^{2l}+iu_{2k}^{2l}\right)e^{i\varphi_{2k+1}},
\end{align}
\end{subequations}
and
\begin{subequations}
  \begin{align}
    u_{2k-1}^{2l+1} & =\frac{1}{\sqrt{2}}\left(u_{2k}^{2l}+iv_{2k}^{2l}\right),\\
    v_{2k-1}^{2l+1} & =\frac{1}{\sqrt{2}}\left(v_{2k-2}^{2l}+iu_{2k-2}^{2l}\right)e^{i\varphi_{2k-1}}.
  \end{align}
\end{subequations}
By combining these equations we reach at:
\begin{subequations}\label{eq:4}
  \begin{align}
    u_{2k}^{2l+2} & = \frac{1}{2}\left(\left(-u_{2k}^{2l}+iv_{2k}^{2l}\right)e^{i\varphi_{2k+1}}+\left(u_{2k+2}^{2l}+iv_{2k+2}^{2l}\right)\right),\\
    v_{2k}^{2l+2} & =\frac{1}{2}\left(\left(-v_{2k}^{2l}+iu_{2k}^{2l}\right)e^{i\varphi_{2k}}+\left(v_{2k-2}^{2l}+iu_{2k-2}^{2l}\right)e^{i(\varphi_{2k-1}+\varphi_{2k})}\right).
  \end{align}
\end{subequations}
Now use the following gauge transformation:
\begin{subequations}\label{eq:gauge}
  \begin{align}
    (-1)^ma_n^m & =-u_{-2k}^{2l},\\
    (-1)^mb_n^m & =+v_{-2k}^{2l}.
  \end{align}
\end{subequations}
In addition we assume:
\begin{equation}\label{eq:phi}
\varphi_n=\varphi_{-2k}=\varphi_{-2k-1}.
\end{equation}
Finally, by replacing Eqs.~(\ref{eq:gauge}) and Eq.~(\ref{eq:phi}) in Eqs.~(\ref{eq:4}) we obtain 
\begin{subequations}\label{eq:meshiteration}
  \begin{align}
    a_{n}^{m+1} & = \frac{1}{2}\left(\left(a_{n}^{m}+ib_{n}^{m}\right)e^{i\varphi_{n}}+\left(-a_{n-1}^{m}+ib_{n-1}^{m}\right)\right),\\
    b_{n}^{m+1} &
    =\frac{1}{2}\left(\left(b_{n}^{m}+ia_{n}^{m}\right)e^{i\varphi_{n}}+\left(-b_{n+1}^{m}+ia_{n+1}^{m}\right)e^{i(\varphi_{n}+\varphi_{n+1})}\right),
  \end{align}
\end{subequations}
which are the evolution equations of optical mesh lattices \cite{Miri2012a}.

For a periodic arrangement of the phase potential, the band structure as well as the corresponding Floquet-Bloch modes are studied in Ref.~\cite{Miri2012a}. In the presence of defects, however, there is not a systematic approach for finding defect states in general. On the other hand, as we will show here, for the specific case where there is a single defect site in a passive empty lattice, an analytical expression for the defect mode can be obtained. For this reason, assume a single defect state imposed on an empty lattice. Therefore, the total phase potential can be written as:

\begin{equation}
\varphi_n=\begin{cases}
 \varphi_d \text{ for } n=0  \\ 
 0 ~~ \text{ for } n\neq 0. 
\end{cases}
\end{equation}
We then assume the following form of solution for a defect mode:
\begin{subequations}
  \begin{align}
    a_n^m& =e^{i\theta m}e^{-\alpha|n|}
    \begin{cases}
      E & n\leq -1  \\
      E \qquad\qquad\text{for} & n=0  \\
      Fe^{-i\varphi_d} & n\geq +1
    \end{cases}\\
    b_n^m& =e^{i\theta m}e^{-\alpha|n|}
    \begin{cases}
      F &  n\leq -1  \\
      E \qquad\qquad\text{for}& n=0  \\
      Ee^{-i\varphi_d} & n\geq +1
    \end{cases}
  \end{align}
\end{subequations}
where $\theta$, $\alpha$, $E$ and $F$ are all unknown constants to be determined. Note, that this solution is propagating along $m$, but is exponentially decaying along both positive and negative values of $n$ and it is therefore very localized to the site $n=0$. In this case just by replacing this simple form of solution into the evolution equations we find the transcendental equations
\begin{subequations}
  \begin{align}
    \cos(\theta_d) & =\frac{1}{2}-\frac{1}{2}\cosh{(\alpha)},\\
    \left(e^{i\varphi_d}+ie^{i\varphi_d}\right)e^{-i\theta_d}+2e^{i\theta_d} & = 1+\left(e^{i\varphi_d}+ie^{i\varphi_d}\right)-e^{-\alpha},
  \end{align}
\end{subequations}
which simultaneously give the confinement factor $\alpha$ and the propagation constant $\theta_d$ for a given defect strength $\varphi_d$. This latter case is plotted in Fig. 2(b) of the main text.

\section{Band structure and Floquet-Bloch modes in optical mesh lattices}

In this section we obtain the band dispersion relation and the continuum of the Floquet-Bloch modes associated with optical mesh lattices. Here we restrict our study to the Hermitian passive lattice described by the evolution Equations~(\ref{eq:meshiteration}) in the main paper. This analysis can be easily extended to the case of $\mathcal{PT}$-symmetric lattices. Consider the following periodic arrangement of the phase modulation $\varphi_n$:
\begin{equation}
\varphi_{n}=
\begin{cases}
+\varphi_{0},~~~n:~ even\\
-\varphi_{0},~~~n:~~ odd
\end{cases}
\end{equation}
In this case Eqs.~(\ref{eq:meshiteration}) reduce to: 
\begin{subequations}\label{eq:reducediteration}
  \begin{align}
a_{n}^{m+1}=\frac{1}{2}\left(\left(a_{n}^{m}+ib_{n}^{m}\right)e^{i\varphi_{n}}+\left(-a_{n-1}^{m}+ib_{n-1}^{m}\right)\right)\\
b_{n}^{m+1}=\frac{1}{2}\left(\left(b_{n}^{m}+ia_{n}^{m}\right)e^{i\varphi_{n}}+\left(-b_{n+1}^{m}+ia_{n+1}^{m}\right)\right)
  \end{align}
\end{subequations}
Now let us assume plane wave solutions of the form:
\begin{equation}\label{eq:planewaves}
\binom {a_{n}^{m}}{b_{n}^{m}}=\binom {A_{n}}{B_{n}}e^{iQn} e^{i\theta m}.
\end{equation}
Since $\varphi_n$ is periodic with a period of $N=2$, the discrete Bloch functions $A_n$ and $B_n$ are also periodic functions of $n$ with the same periodicity, i.e. $A_{n+2}=A_n$ and $B_{n+2}=B_n$. Therefore they can be completely determined by knowing their values in $n=0,1$. Using the ansatz of Eq.~(\ref{eq:planewaves}) in Eqs.~(\ref{eq:reducediteration}) leads to the following eigenvalue problem:
\begin{equation}\label{eq:eigenvalue}
\frac{1}{2}\begin{pmatrix}
e^{i\varphi_0} &ie^{i\varphi_0}  &-e^{-iQ}  &ie^{-iQ} \\ 
ie^{i\varphi_0} &e^{i\varphi_0}  &ie^{iQ}  &-e^{iQ} \\ 
-e^{-iQ} &ie^{-iQ}  &e^{-i\varphi_0}  &ie^{-i\varphi_0} \\ 
ie^{iQ} &-e^{iQ}  &ie^{-i\varphi_0}  &e^{-i\varphi_0} 
\end{pmatrix}
\begin{pmatrix}
A_0\\ 
B_0\\ 
A_1\\ 
B_1
\end{pmatrix}
=e^{i\theta}\begin{pmatrix}
A_0\\ 
B_0\\ 
A_1\\ 
B_1
\end{pmatrix}
\end{equation}
This can be solved to find the dispersion relation:
\begin{equation}
\cos{(2Q)}=8\cos^2{(\theta)}-8\cos{(\varphi_0)}\cos{(\theta)}+4\cos^2{(\varphi_0)}-3
\end{equation}
while the corresponding eigenvectors can be readily obtained.

\section{Passive defects (Fig. 2 of main paper)}

\expfigure{figs/ptarray_phaseshifts_passivedefect}{Optical mesh lattice
  with elemental phase defect $\varphi_d\neq0$ at central positions
  $n={0,1}$ embedded in a passive lattice ($G_p=0$, $G_d=0$) with no
  periodic phase potential ($\varphi_p=0$). Steps $m$ label the rows of
  couplers, while positions $n$ label couplers in horizontal
  direction. The phase shift and the values of gain/loss of a fiber are
  assigned to the position $n$ of the coupler below it. From this
  spatial picture, it becomes clearly visible that every second position
  $n$ is not accessible in each round-trip: Two neighboring couplers have
  a distance of 2 positions $n$ and each row of couplers is shifted by 1
  position $n$ with respect to the rows above and below it. All
  amplitudes at these empty positions $n$ have thus been set to zero in
  all measurements and simulations.}

\expfigure{figs/defect54-messnr_01_suppl}{%
  Data in both loops and comparison to simulations of measurements shown
  in Fig.~2(c). In the main paper, only measurement data recorded in
  the short loop is displayed.}

\expfigure{figs/defect54-messnr_02_suppl}{%
  Data in both loops and comparison to simulations of measurements shown
  in Fig.~2(d).}

\expfigure{figs/defect54-messnr_03_suppl}{%
  Data in both loops and comparison to simulations of measurements shown
  in Fig.~2(e).}

\section{Phase defect in periodic PT lattice (Fig. 3 of main paper)}

\expfigure{figs/ptarray_phaseshifts_ptdefect}{%
  Optical mesh lattice with elemental phase defect $\varphi_d\neq0$ at
  central positions $n={0,1}$ embedded in a periodic PT lattice
  ($G_p\neq0$) with phase potential $\varphi_p\neq0$. The distribution
  of gain/loss has no defect ($G_d=0$).}

\expfigure{figs/defect59-messnr_02_suppl}{%
  Data in both loops and comparison to simulations of measurements shown
  in Fig.~3(c). In the main paper, only measurement data recorded in
  the short loop is displayed.}

\expfigure{figs/defect59-messnr_04_suppl}{%
  Data in both loops and comparison to simulations of measurements shown
  in Fig.~3(d).}

\expfigure{figs/defect59-messnr_06_suppl}{%
  Data in both loops and comparison to simulations of measurements shown
  in Fig.~3(e).}

\section{Gain/loss and phase defect in empty lattice (Figs. 4 and 5 of main paper)}

\expfigure{figs/ptarray_phaseshifts_bicdefect}{%
  Optical mesh lattice with gain/loss defect $G_d\neq0$ and phase defect
  $\varphi_d \neq0$ at 14 positions $n={-6,..,7}$ embedded in a passive
  lattice ($G_p=0$) with no periodic phase potential ($\varphi_p=0$). In
  case of gain/loss defects ($G_d\neq0$), some couplers at the
  boundaries have active (red/blue) fibers on one side and passive
  (black) fibers on the other port in this spatial equivalent picture to
  fulfill the conditions of PT symmetry. To describe this with the
  iteration equations (1) in the main paper, the gain/loss value in the
  short loop is shifted by one position. Therefore, $\widetilde{G}(n+1)$
  is evaluated in the short loop (amplitudes $u_n^m$) while
  $\widetilde{G}(n)$ is evaluated in the long loop (amplitudes
  $v_n^m$).}

\expfigure{figs/bic27-messnr_03_suppl}{%
  Data in both loops and comparison to simulations of measurements shown
  in Fig.~4(c).}

\expfigure{figs/bic27-messnr_06_suppl}{%
  Data in both loops and comparison to simulations of measurements shown
  in Fig.~4(d). A breakdown of localization is observed in experiment after 80 steps in $m$ due to gain saturation of amplifiers.}

\expfigure{figs/bic36-messnr_03_suppl}{%
  Data in both loops and comparison to simulations of measurements shown
  in Figs.~5(b--d).}
  
\section{Experimental setup\label{sec:exp}}

The following sections give a comprehensive description of experimental methods used to obtain the results presented in the publication. For a more accessible and less technical explanation of our setup, please take a look at the Supplemental Material of References \cite{Regensburger2012} and \cite{Regensburger2011}.

\begin{figure}
\centering
\includegraphics[width=0.8\textwidth]{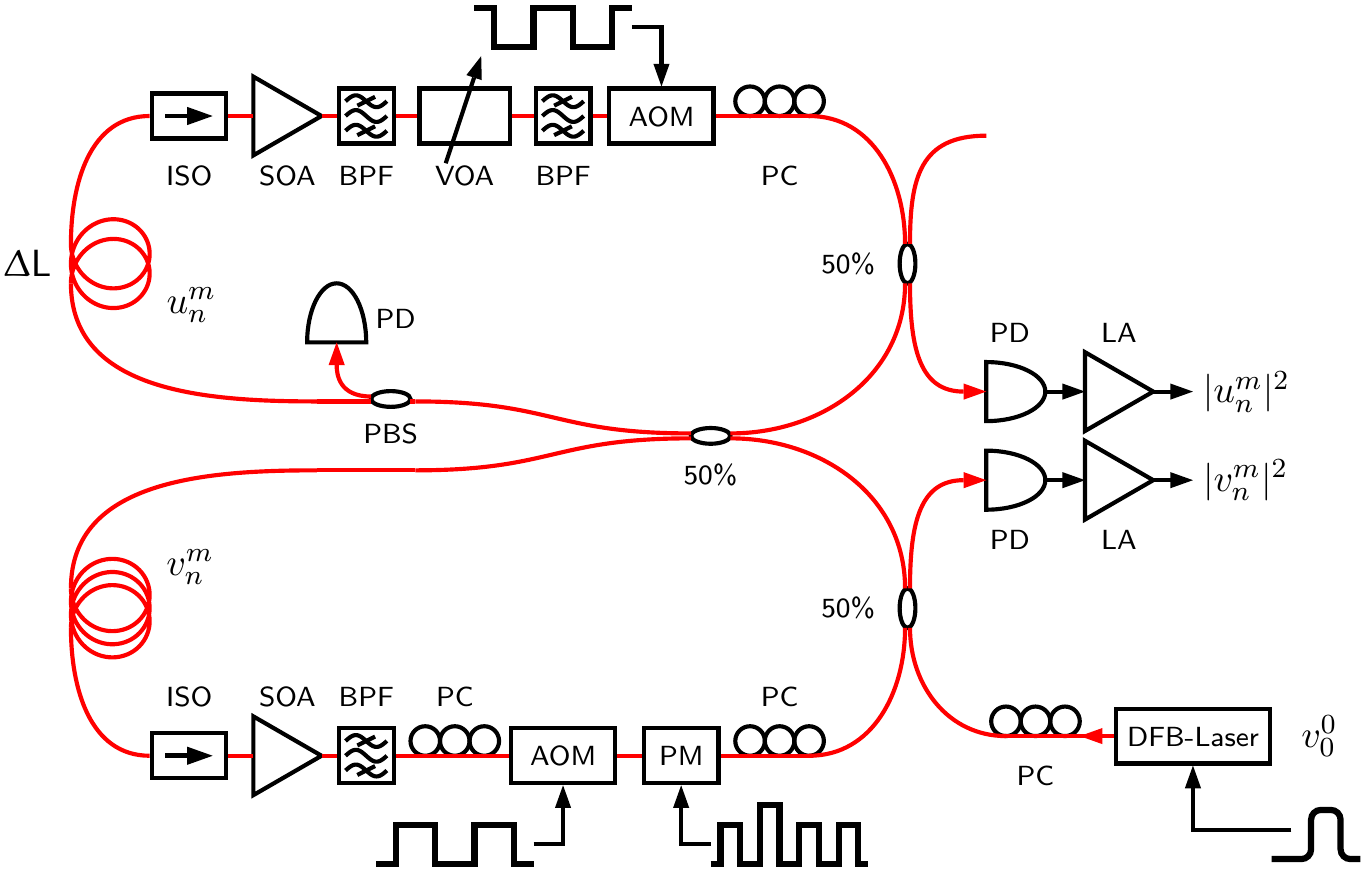}
\caption{Scheme of the fiber setup. Two loops of optical fibers are connected by a central 50\% fiber coupler. $\Delta L$: Length difference between the loops; ISO: Optical Faraday isolator; PC: Polarization controller; BPF: Tunable bandpass filter; SOA: Semiconductor optical amplifier; AOM: Acousto-optic modulator; PM: Phase modulator with integrated polarizer; PD: Photodiode; LA: Logarithmic voltage amplifier.\label{fig:exp}}
\end{figure}

The experimental setup depicted in Fig. \ref{fig:exp} is assembled entirely of standard equipment for C-band optical telecommunication systems. It is based on two loops of standard single-mode fiber that have a measured average length of $L=\SI{23406.6}{\ns} \cdot c_\text{fiber}$, with $c_\text{fiber}$ being the speed of light in the fibers at the signal wavelength. The physically relevant quantity of our measurement is not the actual fiber length, but the optical round-trip times that are used to extract the pulse intensities from the signal (see below). Both loops are connected by a central 50/50 fiber-optical coupler. The length differential $\Delta L=\SI{276.1}{\ns} \cdot c_\text{fiber}$ introduces a transverse coupling between adjacent pulse positions $n$ via time multiplexing (see Section \ref{sec:time-mult} below). An additional 50/50 coupler is used in each loop for signal monitoring. In one loop, this coupler also injects the signal pulse from a DFB laser diode (\textit{SEI SLT5411-CB-F400}) with direct current modulation. It is powered by a \textit{Hytek HY 6510} laser diode driver chip and a \textit{Hytek HY 5610} TEC controller chip. The laser pulses at \SI{1545.3}{\nm} have a close to rectangular shape with duration of \SI{200}{\ns} such that it fits well into the time slot of \SI{276.1}{\ns} fixed by the loop length differential.

Two common telecommunication monitoring photodiodes are connected to the 50/50 monitor couplers in both loops. Their time-varying electrical signals are acquired with a fast digital real-time oscilloscope after being amplified by high-bandwidth logarithmic amplifiers (\textit{Femto HLVA100}). The \SI{100}{\MHz} total bandwidth of the diagnostic system is large enough to resolve the temporal shape and peak intensity of all circulating optical pulses. 

An electro-optical lithium-niobate phase modulator (\textit{Covega LN65S}) is driven by a \SI{100}{\MHz} arbitrary waveform generator (\textit{Tektronix AFG3102}) to apply the real-valued potential function $\varphi (n,m)$ on the pulse phases in the long loop. This way, a computer-generated time-dependent voltage signal is effectively translated into the temporal equivalent of a transverse distribution of the refractive index in the optical mesh lattice (see Section \ref{sec:ptsymm} and Fig. \ref{fig:timing-scheme}). The voltage function is programmed such that it also contains the phase defects required for studying light localization around defects in passive and non-Hermitian optical mesh lattices. It is stored in the integrated memory of the waveform generator and repeated at a rate of \SI{25}{\Hz} at which the whole experiment is operated. All other electrical signals necessary to control the setup operation are also synthesized by waveform generators (\textit{Agilent 33522A}). 

Two semiconductor optical amplifiers (\textit{Finisar G111}) do not only counterbalance all round-trip losses stemming from component losses and leakage at the monitor couplers, but also provide a constant amount of excess gain. With the acousto-optic modulators (AOM, \textit{Neos 26027} operated in zeroth diffraction order) in their transparent state, the variable attenuators are used to adjust the round-trip losses such that all pulse intensities are multiplied by a residual gain factor of $1.46$. Depending on the gain/loss modulation parameters, the AOMs attenuate the signals in the loops to have a certain amount of net gain or net loss. This enables the temporal implementation of arbitrary distributions of optical amplification and attenuation for each pulse that are prepared in the way necessary to realize $\mathcal{PT}$ symmetry in the structure. At the boundaries between two different steps $m$ and $m+1$, the AOMs are set to a highly absorptive state to prevent uncontrolled noise build-up and signal cross-talk between two steps (see Section \ref{sec:ptsymm} and Fig. \ref{fig:timing-scheme}). Between two successive measurements, the AOMs were set to maximum absorption to erase all circulating optical signals from the loops.

Several additional passive optical fiber components are used for polarization and gain control as well as for noise filtering.  One fiber-coupled optical Faraday isolator is present in each loop to filter out counter-propagating noise and back reflections. In addition, a single bandpass filter in the long loop (\textit{DiCon TF-1565-0.8-FC/APC-0.3-1}) and a combination of two bandpass filters in the short loop (\textit{JDS Uniphase VCF050-Z001} and \textit{JDS Fitel TB1500B}) suppress excess noise due to amplified spontaneous emission of the amplifiers. Several fiber-quenching polarization controllers and another monitoring photodiode at the rejection port of a fiber-coupled polarizing beam splitter in the short loop are used to adjust and monitor the state of polarization. Additionally, the phase modulator has an integrated polarizer that preserves a clean state of polarization in the long loop. The static round-trip losses of the short loop are tuned with a precision variable optical attenuator (\textit{JDS Fitel}). In the other loop, a polarization controller in front of the polarizer of the phase modulator serves the same purpose. 

Directly after each signal acquisition, which is averaged over a sufficient number of realizations, the residual amplifier noise is recorded in a dark frame that is realized under exactly the same conditions except for the laser source being switched off. It is averaged the same way as the signal frame and subsequently subtracted point by point. Afterwards, the intensities of pulse peaks are extracted from the data (see Section \ref{sec:time-mult}). Here, an additional arithmetic mean along several points of the temporal waveform is applied in the evaluation procedure. 

All measurements within the same figure in the Letter were recorded in a subsequent measurement series where the parameter variation was realized by a change of the modulation waveform only. No manual intervention e.g. to adjust the static net gain of the setup was performed in between the measurements in a series. 

\section{Time multiplexing\label{sec:time-mult}}

\begin{figure}\centering
\includegraphics[width=0.8\textwidth]{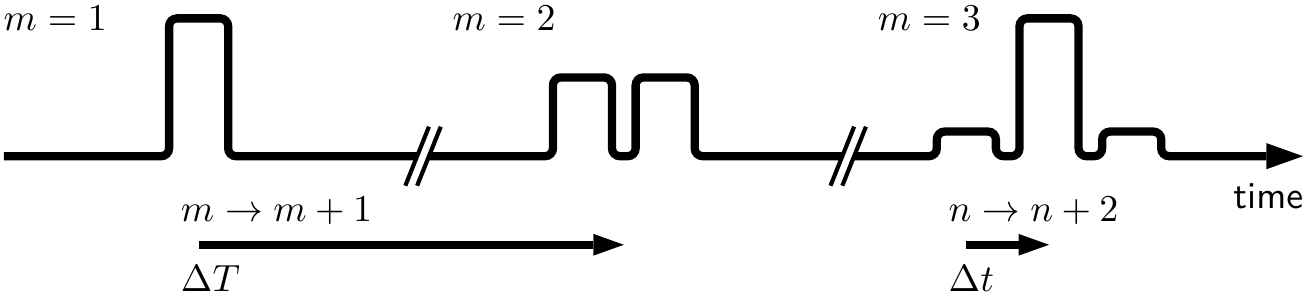}
\caption{Time multiplexing. Schematic drawing of the time trace of the optical power $U(t)$ as recorded by the photodiode attached to the short loop. Step number $m$ and position $n$ are indicated. Source: Ref. \cite{Regensburger2012}.\label{fig:timeline}}
\end{figure}

In our experimental approach, time multiplexing is applied to realize a $\mathcal{PT}$-symmetric optical lattice in the temporal domain. Figure \ref{fig:timeline} shows a sketch of the signal waveform $U(t)$ as detected by the photodiode attached to the monitor coupler in the short loop. At first, the single pulse that is initially inserted into the long loop gets detected after having completed one trip around the short loop (step $m=1$). One loop round-trip later (after a time interval $\Delta T=L/c_\text{fiber}$ with $L$ being the average length of the two loops), two neighboring pulses with a significantly shorter relative time offset $\Delta t=|\Delta L|/c_\text{fiber} \ll T$ are recorded by the photodiode. The pulse that arrives first went around the short loop two times, while the second pulse first travelled through the long and afterwards through the short loop. After the pulses have circled the loops one more time ($m=3$), a sequence of three pulses is observed, with the central one being subject to wave interference. The coherent superposition arises because this temporal position can be reached via two alternate paths through the two-loop network (long--short or short--long), see Fig.~1(b) of the main paper. 

The pulse intensities in the acquired signal $U(t)$ are then extracted to the discrete $m \times n$ coordinate system of the mesh lattice via $\left|u_n^m \right|^2=U\left(m \Delta T + \frac{n}{2} \Delta t \right)$ with the above-mentioned read-out timings $\Delta T$ and $\Delta t$. Following this notation, only every second position $n$ can be physically accessed by the pulses. This simplifies the analytical treatment and results in a symmetric behavior of the system. The initial pulse starts at step $m = 0$ and the central position $n = 0$. In all round-trips where the step $m$ is an even number ($m = 0, 2, 4,\ldots$), only even positions $n$ are physically relevant. In all other round-trips ($m=1,3,5,\ldots$) only odd positions $n$ can be reached physically. Signals at physically non-relevant positions are set to zero. Finally, the extracted data is visualized on the discrete 2D $m \times n$ grid in a logarithmic color scale as indicated next to all plots. 

To conduct the simulations of signal propagation shown in this Supplemental Material, Eq. (1) of the main paper is evaluated numerically with the initial condition of a single pulse having amplitude $1$ in the long loop and the parameters stated in the respective figure captions. 

\section{Experimental implementation of parity-time symmetry\label{sec:ptsymm}}

\begin{figure}\centering
\includegraphics[width=0.8\textwidth]{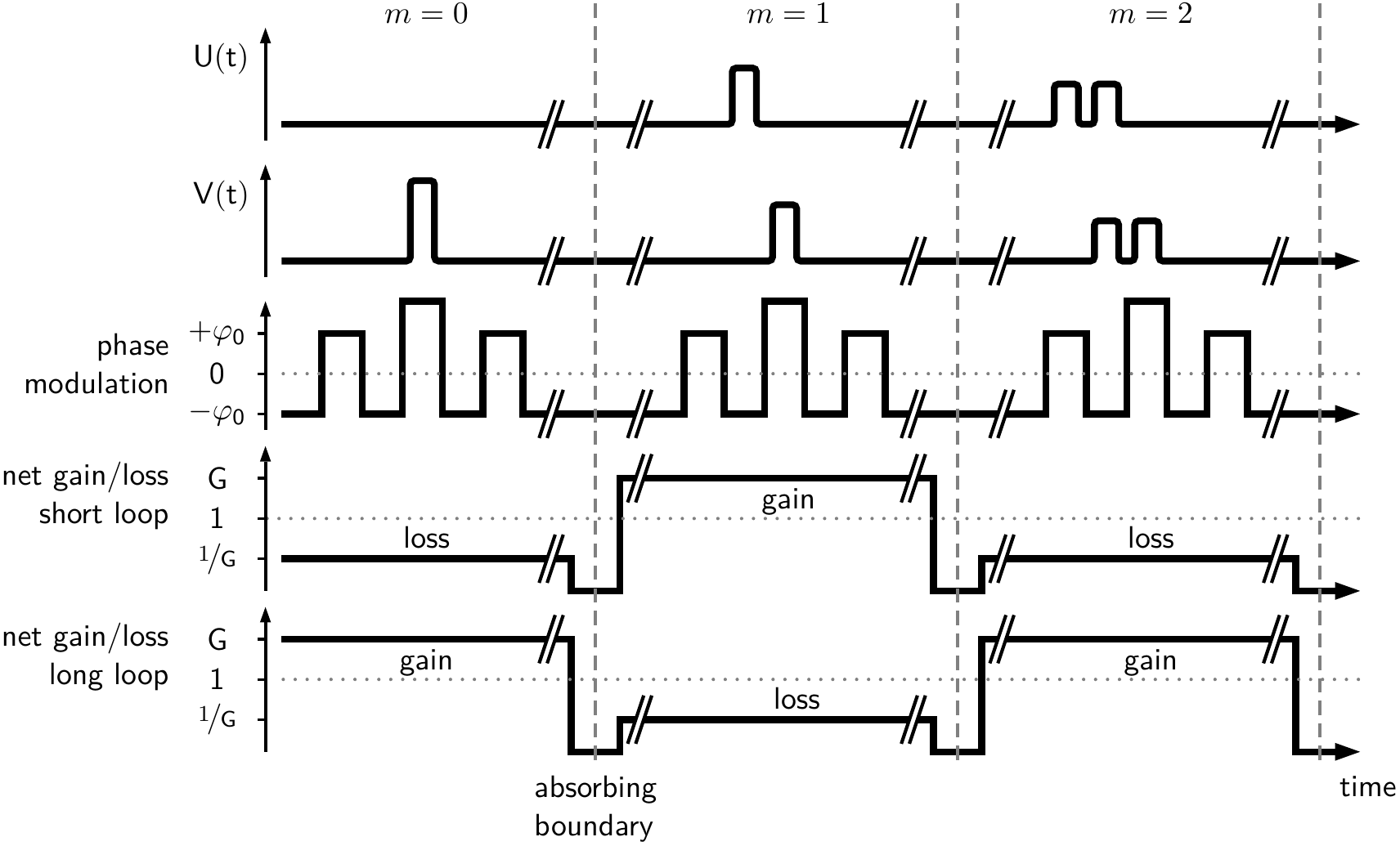}
\caption{Timing scheme of modulation to impose a $\mathcal{PT}$-symmetric optical potential. Simplified traces are displayed; pulse power levels do not account for amplification or attenuation of signal. The phase modulation contains a central defect. Source: Adapted from Ref. \cite{Regensburger2012}.\label{fig:timing-scheme}}
\end{figure}

For the realization of time-dependent gain/loss in the system, the AOMs in both loops are modulated to switch the net gain or net loss between pulses thus enabling a completely arbitrary distribution of the imaginary optical potential throughout all lattice points of the time-multiplexed optical mesh.

The operation of the experiment is based on a computer-programmed modulation scheme of the loop losses by AOMs and of the signal phase by the phase modulator, as displayed in Fig. \ref{fig:timing-scheme}. Here, $U(t)$ and $V(t)$ illustrate the temporal traces detected by the monitoring photodiodes in the short and long loops. The phase modulator applies the same phase function $\varphi(n)$ in every loop round-trip. In the case shown, it contains a central phase defect around which localized modes will arise during propagation. Finally, the AOM modulation signals, used to obtain net gain and net loss which obeys parity-time symmetry, are indicated for both loops. As required by $\mathcal{PT}$ symmetry, the two loops are repeatedly switched between gain and loss at every loop round-trip.